\documentclass{aa}  
\usepackage{lscape}
\usepackage{longtable,threeparttable}
\usepackage{graphicx}
\usepackage{txfonts}
\usepackage[switch]{lineno}
\usepackage{caption}
\usepackage{subcaption}
\usepackage{hyperref}
\hypersetup{
	unicode=false,
	pdftoolbar=true,
	pdfmenubar=true,
	pdffitwindow=false,
	pdfstartview={FitH},
	pdftitle={My title},
	pdfauthor={Author},
	pdfsubject={Subject},
	pdfcreator={Creator},
	pdfproducer={Producer},
	pdfkeywords={keyword1} {key2} {key3},
	pdfnewwindow=true,
	colorlinks=true,
	linkcolor=red,
	citecolor=blue,
	filecolor=magenta,
	urlcolor=cyan
}
\newcommand{\tess}{{\em TESS}}
\newcommand{\kepler}{{\em Kepler}}
\newcommand{\ktwo}{{\em K2}}
\newcommand{\deltaphi}{{$\Delta\phi_{1-0}$}}
\newcommand{\ampratio}{$a_1/a_0$}
\newcommand{\lcurve}{\texttt{lcurve}}
\begin{document} 

%%%%%%%%%%%%%%%%%%%%%%%%%%%%%
%%%       META DATA       %%%
%%%%%%%%%%%%%%%%%%%%%%%%%%%%%
   \title{Hot subdwarfs in close binaries observed from space III: \\Reflection effect asymmetry induced by relativistic beaming}
   
    \titlerunning{Reflection Effect Asymmetry from Doppler Beaming}
   \author{B.~N. Barlow \inst{1,2}\thanks{E-mail: bbarlow@highpoint.edu}
          \and T. Kupfer\inst{3,4}  
          \and B.A. Smith\inst{1}
          \and V. Schaffenroth\inst{5,6}
          \and I. Parker\inst{1}
          }
\authorrunning{Barlow et al.}
        \institute{Department of Physics and Astronomy, High Point University, High Point, NC 27268, USA
        \and
        Department of Physics and Astronomy, University of North Carolina at Chapel Hill, Chapel Hill, NC 27599, USA
        \and
        Hamburger Sternwarte, University of Hamburg, Gojenbergsweg 112, D-21029 Hamburg, Germany
        \and
        Department of Physics and Astronomy, Texas Tech University, P.O. Box 41051, Lubbock, TX 79409, USA
        \and
        Thüringer Landessternwarte Tautenburg, Sternwarte 5, D-07778 Tautenburg, Germany
        \and
       Institute for Physics and Astronomy, University of Potsdam, Karl-Liebknecht-Str. 24/25, 14476 Potsdam, Germany
             }
   \date{Received: 28 October 2023 / Accepted: 20 February 2024}

%%%%%%%%%%%%%%%%%%%%%%%%%%%%%%%%
%%%       INTRODUCTION       %%%
%%%%%%%%%%%%%%%%%%%%%%%%%%%%%%%%
% \abstract{}{}{}{}{} 
% 5 {} token are mandatory
  \abstract{Detailed studies of hot subdwarf B stars with red dwarf or brown dwarf companions can shed light on the effects of binarity on late stellar evolution. % and help determine whether there is a lower mass limit for stripping a red giant and surviving common envelope evolution. 
  Such systems exhibit a strong, quasi-sinusoidal reflection effect due to irradiation of the cool companion, and some even show primary and secondary eclipses. %Before space-based photometric surveys, light curves of sdB+dM/BD binaries were rarely subjected to Fourier analyses due to the limited number of orbits they were observed and complicated window functions from daytime and seasonal observing gaps. Continuous 27+ day observations from \tess\ make such an analysis fruitful for a large number of objects for the first time. 
  Here we compute Fourier transforms of \tess\ light curves of sdB+dM/BD binaries and investigate correlations between the relative amplitudes and phases of their harmonics and system parameters. We show that the reflection effect shape strongly depends on the orbital inclination, with nearly face-on systems having much more sinusoidal shapes than nearly edge-on systems. This information is encoded by the relative strength of the first harmonic in the Fourier transform. By comparing observations of solved systems to synthetic light curves generated by \lcurve , we find that the inclination of non-eclipsing systems with high S/N light curves can be determined to within $\approx$10$^{\circ}$ simply by measuring their orbital periods and first harmonic strengths. We also discover a slight asymmetry in the reflection effect shape of sdB+dM/BD binaries using the relative phase of the first harmonic. From our analysis of synthetic light curves, we conclude the asymmetry results from relativistic beaming of both stellar components. This marks the first time Doppler beaming has been detected in sdB+dM/BD systems. Although advanced modeling is necessary to quantify the effects of secondary parameters like limb darkening, the temperature ratio, and the radius ratio on the reflection effect shape, our pilot study demonstrates that it might be possible to extract both the inclination angle and cool companion velocity from the light curves of non-eclipsing systems.}


   \maketitle

%%%%%%%%%%%%%%%%%%%%%%%%%%%%%%%%
%%%       INTRODUCTION       %%%
%%%%%%%%%%%%%%%%%%%%%%%%%%%%%%%%
\section{Introduction}
\label{sec:intro}

Hot subdwarf B (sdB) stars are a natural choice for studying the effects of stellar and substellar companions on late stellar evolution. It is widely accepted they are the progeny of red giant branch (RGB) stars that were stripped of their outer H envelopes while ascending the RGB. This process reveals a hot, dense He core surrounded by a thin H atmosphere: a hot subdwarf. They are located in the Hertzsprung--Russell diagram between the main sequence and white dwarf cooling tracks near or on the so--called extreme horizontal branch \citep{greenstein74}. Models show many sdBs are core He--burning stars surrounded by a thin H atmosphere, with canonical masses near 0.5 M$_{\sun}$ and radii around 0.2 R$_{\sun}$ \citep{heber86,saffer94}. Due to a dearth of outer H, they will not ascend the asymptotic giant branch, as do most horizontal branch stars.  Instead, they will exhaust all of the He in their cores after $\sim$100 Myr and directly enter the white dwarf cooling sequence \citep{dorman93}. Other non He-fusing sdB stars might instead be pre-extremely low mass white dwarfs (pre-ELM WDs) on their way to the WD cooling sequence (e.g., \citealt{ratzloff19})

Binarity is the most widely-accepted mechanism for the formation of hot subdwarf B stars, with several such channels having been proposed that invoke Roche lobe overflow (RLOF) and common envelope (CE) evolution processes (e.g., \citealt{han02,han03}). Observations seem to support this hypothesis, with recent studies even arguing that binary interactions are {\em always} required for the formation of a hot subdwarf \citep{pelisoli20}. Around one--third of all sdBs are in close binaries with M dwarf (dM), brown dwarf (BD), or white dwarf (WD) companions and have orbital periods ranging from a few hours to several days \citep{maxted01,kupfer15a, schaffenroth19}. Since their separation distances are much smaller than the diameter of the red giant progenitors, they must have gone through a common envelope phase, and consequently, the companion must have been responsible for the strong mass loss required to create the sdB. Thus, detailed studies of the mass and orbital period distributions may shed light on their evolutionary histories and help tune parameters in binary population synthesis models. 

Hot subdwarf binaries with cool dM/BD companions are relatively easy to find from the strong photometric variations they exhibit. The most striking feature of their light curves is the so--called reflection effect. UV flux from the hot subdwarf irradiates the surface of the cool companion facing it, and as this heated face swings in and out of view from Earth, it results in a quasi--sinusoidal light curve variation at the orbital period of the binary. The large surface temperature difference between the stars results in a reflection effect amplitude that can approach 10-20\% in the optical. For systems with hot white dwarf primaries, the reflection effect can be even stronger. Many sdB+dM/BD binaries also show primary and secondary eclipses, in which case they are referred to as `HW Vir' binaries. Observed eclipse depths range from grazing eclipses (e.g., \citealt{schaffenroth13}) to what are essentially total eclipses (e.g., \citealt{corcoran21}). Despite their being single--lined spectroscopic binaries, masses in eclipsing systems can be determined by combining spectroscopic results or SED fitting for the sdB primary ($K_{\rm sdB}$, log $g$, and $T_{\rm eff}$) with detailed light curve modeling (e.g., \citealt{schaffenroth19,schaffenroth21}).

The Transiting Exoplanet Survey Satellite (\tess) has provided the first expansive, high signal-to-noise (S/N) data set of post-CE sdB+dM/BD binaries. In \citet{schaffenroth22} (hereafter, Paper I) we conducted a search for new variable hot subdwarfs in \tess\ and found 85 new systems showing a reflection effect. The incredibly high quality of the \tess\ photometry not only permitted light curve modeling for the eclipsing (HW Vir) binaries, but also for the non-eclipsing ones for the first time. While eclipses had traditionally been required to constrain the inclination angle, the \tess\ data revealed that the shape of the reflection effect also encodes the system inclination: systems with low inclination have much more sinusoidal reflection effect shapes than those that are nearly eclipsing. The complete set of light curves, along with full details regarding our modeling and analysis methods, are presented
in \citet{schaffenroth23} (herefter, Paper II). The \tess\ data allowed us to derive the first orbital period and mass distributions for sdB+dM/BD binaries selected from light variations instead of radial velocity variations. We find no systems with periods shorter than 1.5 hrs. Below this, companions of any type would exceed their Roche radii around a canonical-mass sdB (see Fig 14 of \citealt{schaffenroth19}). Additionally, we find that the cool, low-mass companions cluster around the hydrogen-burning limit.

The light curves of reflection effect sdB+dM/BD binaries have not traditionally been subjected to detailed Fourier analyses due to the limited number of orbits for which they are observed or complicated window functions from gaps when observed from the ground. The high quality and relatively continuous nature of \tess\ photometry has made such analyses fruitful only recently. \citet{barlow22}, for example, demonstrated that the relative amplitudes and phases of harmonics in the discrete Fourier transforms (DFTs) of \tess\ light curves can be used to classify certain types of hot subdwarf variables, including those with cool dM/BD companions. 

Here we explore simple Fourier analyses of \tess\ light curves of sdB+dM/BD binaries. In Section \ref{sec:phot} we discuss light curves from \tess\ and the creation of synthetic light curves to test our analyses and interpret our results. In Section \ref{sec:methods} we present the details of our Fourier analyses, along with a simple mathematical expression for the reflection effect shape. In Section \ref{sec:inclination} we show that the inclination angle in non-eclipsing sdB+dM/BD binaries can be constrained by the relative strength of the first harmonic. In Section \ref{sec:asymmetry} we present evidence that the reflection effect is asymmetric and argue that this can be attributed to Doppler beaming of both the sdB and cool companion. We summarize and discuss the implications of our results in Section \ref{sec:summary}.\\

%Despite the ease with which they can be discovered, there are some barriers to determining the masses and other system parameters for sdB+dM/BD systems. First, only around . First, they are single--lined spectroscopic binaries in which the primary outshines the secondary by a factor $>$1000 in the optical. Radial velocities can thus be measured only for the primary.\footnote{In rare cases, faint emission lines from the irradiated side of the companion have been detected.} \citep{vuckovic16}. 

%We recently started the Eclipsing Reflection Effect Binaries from the OGLE Survey (EREBOS) project to obtain follow--up spectroscopy and determine the orbital velocities and atmospheric parameters of the sdB primaries in these newly--discovered binaries. Spectroscopic results can be combined with light curve modeling to determine the component masses in these systems (e.g., \citealt{bar13}). The overarching objective of the EREBOS project is to solve for the masses in each of these systems in order to determine whether there is a lower mass limit at which the companions do not survive the red giant engulfment but instead evaporate.  Full details regarding the EREBOS project may be found in the forthcoming paper by Schaffenroth et al. (in prep). 

%%%%%%%%%%%%%%%%%%%%%%%%%%%%%%
%%%       PHOTOMETRY       %%%
%%%%%%%%%%%%%%%%%%%%%%%%%%%%%%
\vspace{0.1cm}
\section{Photometric Data}
\label{sec:phot}

%%%%%%%%%%%%%%%%%%%%%%%%%%%%%%%%%%%%%%%%%%%%%
\subsection{Observed Light Curves from \tess}
\label{sec:phot_tess}

For the past several years, the \tess\ mission has been collecting high signal--to--noise (S/N) time--series photometry of millions of objects across the entire sky \citep{ric15}. Observations are carried out with four small telescopes that together image a 24$^{\circ}$x96$^{\circ}$ strip repeatedly over 27--d long sectors. Full--frame images (FFIs) were taken every 30 min (primary mission), 10 min (first extension), or 200 s (second extension) for all stars falling in the field of view, downloaded to Earth, and made publicly available. Some stars were also chosen for 2--min or 20--sec cadence observations, primarily through the \tess\ Guest Investigator (GI) program. Although stars at low eclpitic latitudes might only be observed for one sector at a time every couple of years, stars at higher ecliptic latitudes in the continuous viewing  zone are observed for numerous sectors in a row. 

Several thousand known and candidate hot subdwarf stars from \citet{gei19} and \citet{cul22} have been observed with 2--min and/or 20--sec cadence to date, primarily through the \tess\ Asteroseismic Consortium (TASC) and GI programs G022141, G03221, G04091, \& G04116. As discussed in Paper I, \tess\ has observed several known and new hot subdwarf in close, post--CE binaries with low--mass main sequence dM or BD companions. We used the the Python package \texttt{lightkurve} \citep{lightkurve18} to download calibrated light curves for all compact sdB+dM/BD binaries observed in Sectors 1--36 from the Mikulski Archive for Space Telescopes\footnote{https://archive.stsci.edu/missions-and-data/tess} (MAST). The data were automatically
reduced and corrected for instrumental systematics using the
\tess\ data processing pipeline\footnote{https://heasarc.gsfc.nasa.gov/docs/tess/pipeline.html} \citep{jen16}. For the flux we adopt the presearch data--conditioning \texttt{PDCSAP\_FLUX} values, which are aperture photometry values
corrected for systematic trends. If a star was observed in multiple sectors, we combined all observations together into a single light curve after removing any systematic offsets in the average flux from sector to sector. Since the large 21'' pixels can lead to contamination of the light curve by nearby, unresolved stars, we recorded the \texttt{CROWDSAP} value for each target, which measured the ratio of target flux to total flux falling in the \tess\ aperture.

In order to verify and understand the results of the Fourier analysis presented in Section \ref{sec:methods}, we generated a grid of synthetic light curves of sdB+dM/BD binaries using the code \lcurve\ \citep{cop10}. \lcurve\ uses grids of points to model the two stars in a binary. The shape of the stars in the binary is set by a Roche potential. The orbit is assumed to be circular and the rotation periods of the stars are synchronized to the orbital period. The flux that each point on the grid emits is calculated by assuming a blackbody of a certain temperature at the bandpass wavelength, corrected for limb darkening, gravity darkening, Doppler beaming and the reflection effect. To model the reflection effect, \lcurve\ uses the absorb coefficient which defines the fraction of irradiating flux from the sdB absorbed and reemitted by the low mass companion. sdB$+$dM/BD binaries typically require very high or even non-physical albedo values to model the reflection effect strength --- sometimes as high as 200\%\ (e.g., \citealt{for2010}). This arises from light curve modeling codes struggling to model the irradiation strength properly in extreme temperature ratio systems like ours, in which the M dwarf photosphere reprocesses most of the incident UV flux to much lower frequencies. In our models, we assume that 100\%\ of the flux is absorbed and reemitted for all binaries. The passband-specific beaming parameter $B$ ($F_\lambda = F_{0,\lambda} \lbrack 1 - B \frac{v_r}{c}\rbrack$, see \citealt{blo11}) was calculated following the approximation from \citet{loe03}. The beaming factor arises from three main contributions. First, the increased rate of photons from an approaching source contributes $+$1. Second, the number of photons observed from an approaching source also increases due to aberration, which adds $+$2 to the factor (a result of the squared relation between the solid and normal angles). Finally, the contribution from the Doppler shift affects the beaming factor. This contribution can be either positive or negative depending on the temperature of the object. In the case of an sdB, with a spectrum peaking in the UV, the flux of an approaching sdB will decrease when observed at optical wavelengths (in the longer-wavelength tail of the spectrum) due to the entire spectrum shifting towards the blue. This gives rise to a negative beaming factor contribution. In the case of a cooler dM/BD, with a significantly redder spectrum, the flux in the optical will increase, giving rise to a positive beaming factor contribution.   Additional details on the Doppler beaming factor contributions and their values may be found in \citet{blo11}. For our synthetic systems, we adopt overall beaming parameters of $B=1.34$ for the sdB and $B=5.73$ for the cool companion. The passband specific gravity-darkening ($\beta$) and limb-darkening ($a_\mathrm{1}$, $a_\mathrm{2}$, $a_\mathrm{3}$, $a_\mathrm{4}$) were fixed to the theoretical values specific for the \tess\ filter taken from \citet{claret17} for $T_1=29,000$\,K and $\log(g)=5.00$ and solar metallicity. We used $\mathrm{\beta}=0.42$, $\mathrm{a_1}=0.8547$,  $\mathrm{a_2}=-1.0781$, $\mathrm{a_3}=0.9184$, and $\mathrm{a_4}=-0.3025$ for the sdB and $\mathrm{\beta}=0.31$, $\mathrm{a_1}=0.6274$,  $\mathrm{a_2}=0.2536$, $\mathrm{a_3}=-0.0939$, and $\mathrm{a_4}=-0.0140$ for the cool companion.

Our goal was {\em not} to generate a model specific to each individual target, but instead to understand basic trends in the shape of the reflection effect of sdB+dM/BD binaries. To this end, we created a single pair of stars and varied only their orbital periods and inclinations. We fixed the hot subdwarf mass to the canonical value $M_1 = 0.47 M_{\odot}$ and the companion mass to $M_2 = 0.2 M_{\odot}$. For the hot subdwarf, a temperature of $T_1 = 29000 K$ and radius of $R_1 = 0.2 R_{\odot}$ were used. For the cool companion, we adopt $R_2 = 0.22 R_{\odot}$ and $T_2 = 3200 K$, based off of the relations presented by \citet{bar15}. The masses and radii were chosen to match the observed distribution peaks for solved sdB+dM/BD binaries shown in Paper I. We note that the radii quoted are {\em volumetric} radii. \lcurve\ uses the radius measured along the line of centres towards the companion star. We corrected the volumentric radii and found that even at short orbital periods the correction is less than a few percent.

We generated models of the above stars for a range of periods from 2.5 hr (0.1042 d) up to 15 hr (0.625 d), in increments of 30 min (0.0208 d). Once again, this period range was chosen to cover the observed period distribution of solved binaries in Paper I. We note that orbital periods shorter than 0.1042 d are known, but systems with our fixed combination of masses and radii would fill their Roche lobes and begin mass transfer at such periods. For each fixed period, we generated models over a range of inclinations from 5$^{\circ}$ up to 90$^{\circ}$, in increments of 5$^{\circ}$. Thus, both eclipsing (HW Vir) and non--eclipsing (reflection effect only) light curves were generated. To simulate the TESS cadence, model light curvse were created using a 2\,min cadence with a 28\,day baseline. 

%Gravity and limb darkening coefficients were fixed to the values in \citet{cla11} closest to the atmospheric parameter of this binary for the I band filters, which are closest to the \tess\ filter. We modelled relativistic beaming ($F$) as shown in \citet{blo11} and calculated the beaming parameters by assuming a blackbody spectrum and using the effective wavelength of the \tess\ filter, for which we find F=\#. Synthetic light curves were sampled once every 2 min for a period of 27 days, in order to emulate the \tess\ observations.

%%%%%%%%%%%%%%%%%%%%%%%%%%%%%%%%%%%
%%%       FOURIER SERIES       %%%
%%%%%%%%%%%%%%%%%%%%%%%%%%%%%%%%%%%
\vspace{1.75cm}
\section{Fourier Series Analysis}
\label{sec:methods}

%%%%%%%%%%%%%%%%%%%%%%%%%%%%%%%%%%%%%%%%%%%%%%%%%%%%%%%%
\subsection{Synthetic Light Curves from \lcurve}
\label{sec:phot_lcurve}

%%%%%%%%%%%%%%%%%
\begin{figure*}[t]
    \centering
    \includegraphics[width=0.495\textwidth]{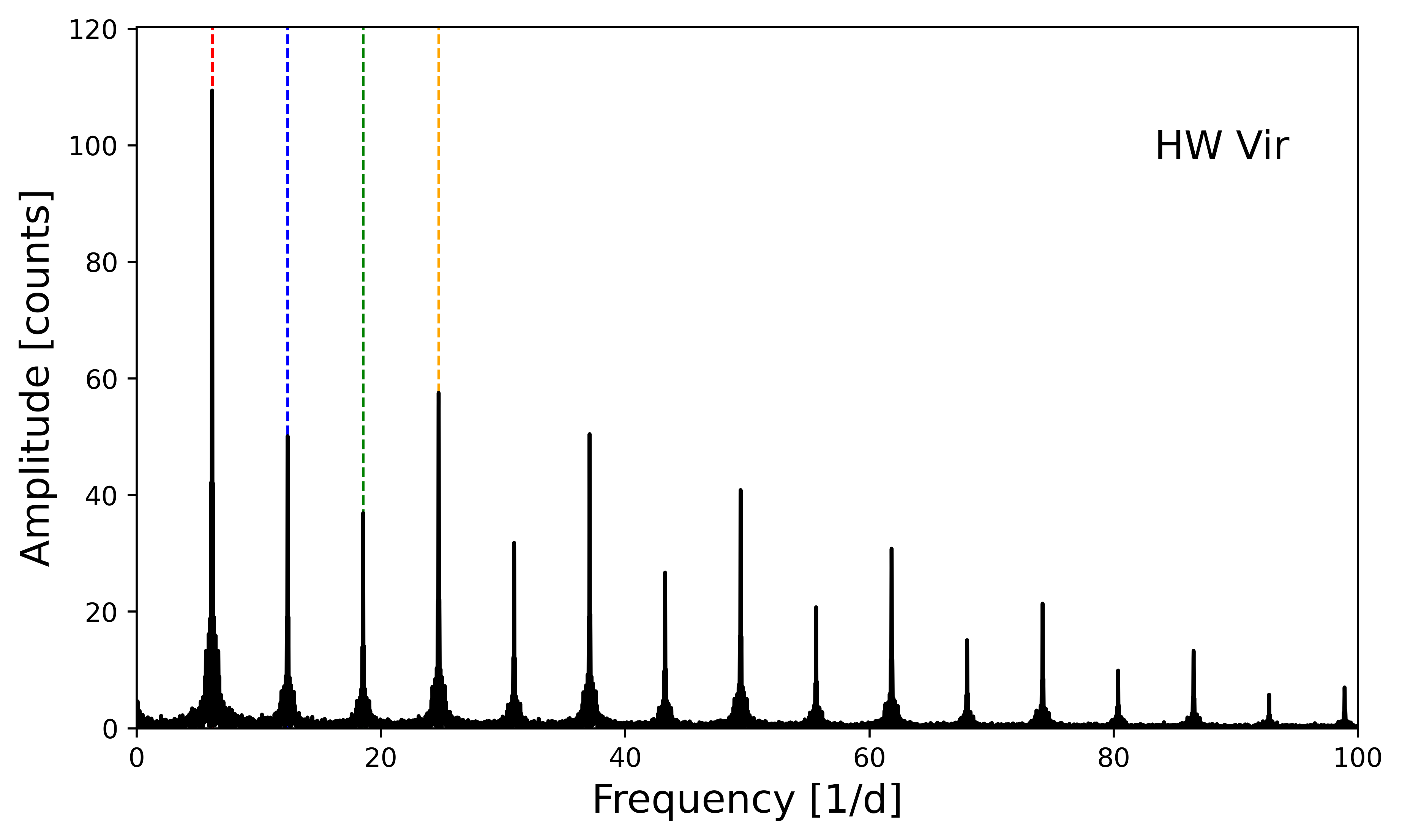} %width 85
  \includegraphics[width=0.495\textwidth]{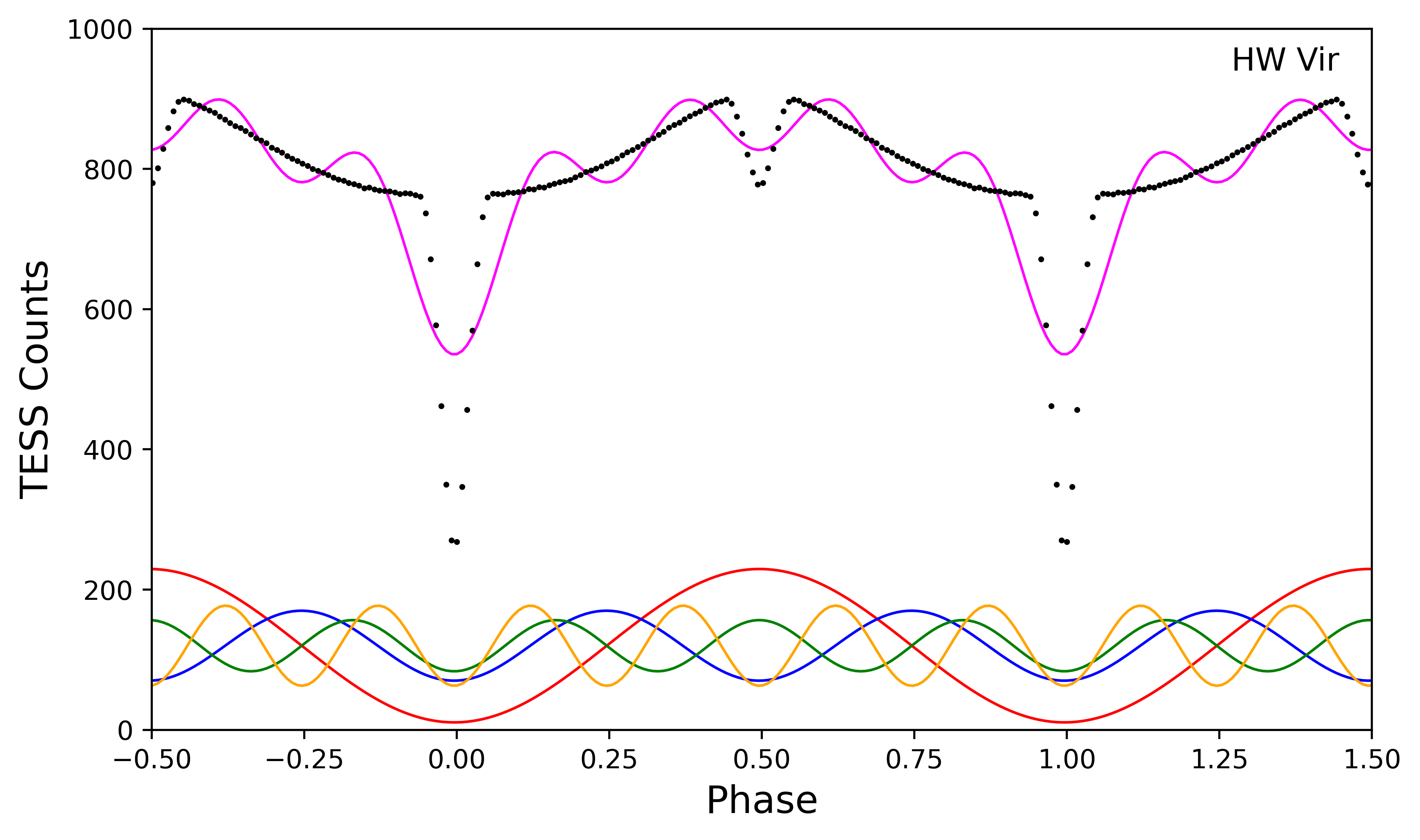}
    \includegraphics[width=0.495\textwidth]{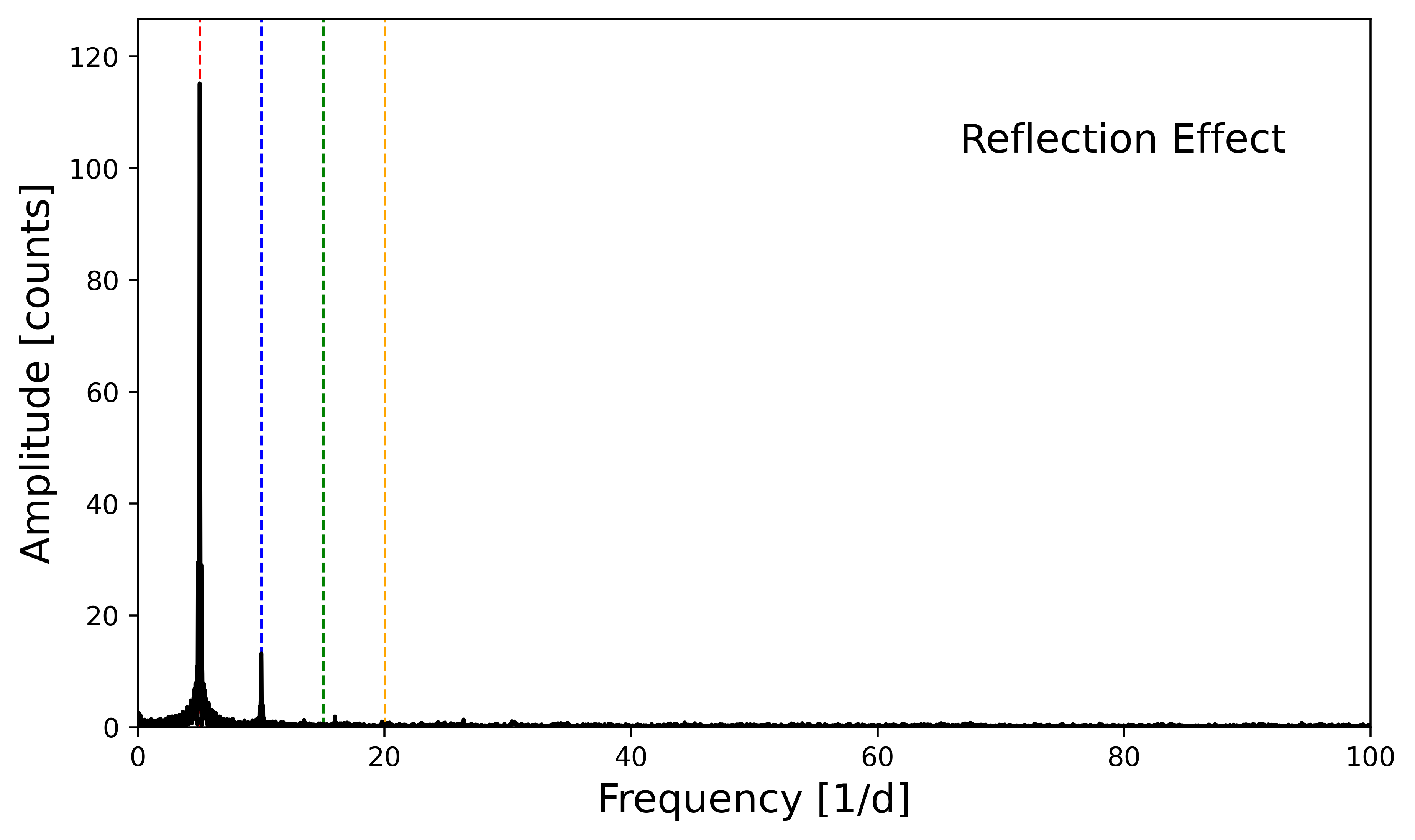}
      \includegraphics[width=0.495\textwidth]{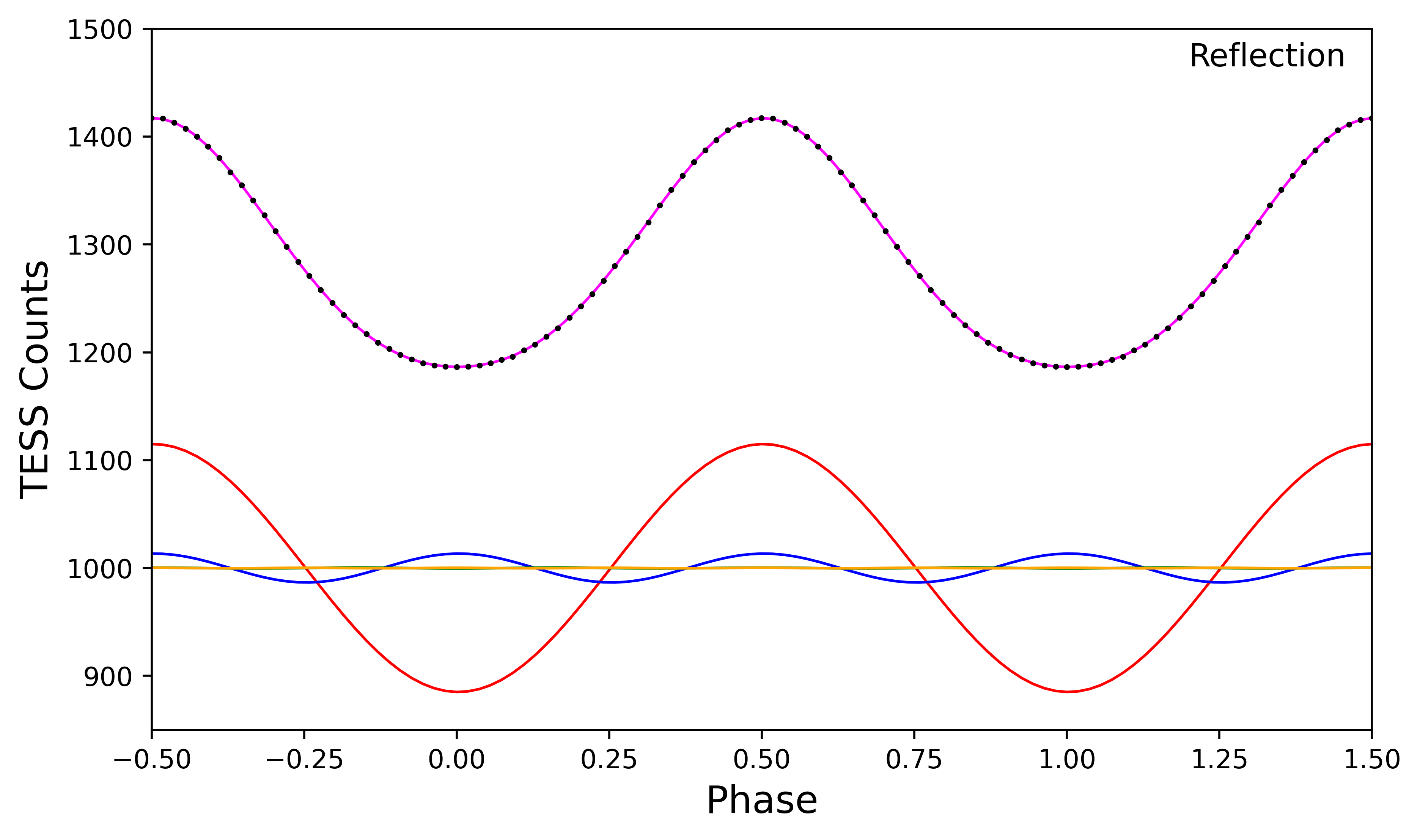}
    \caption{Discrete Fourier transforms (left panels) of example \tess\ light curves (right panels) of eclipsing HW Vir (top) and non-eclipsing (bottom) sdB+dM/BD binaries. Each light curve shape can be reproduced by the summation of a fundamental signal and harmonics with the correct relative phases and amplitudes. The fundamental (red) along with the first (blue), second (green), and third (orange) harmonics are marked in the Fourier transforms and represented by their individual sinusoids below the \tess\ light curves. The magenta lines show their superposition.}
    \label{fig:ft_examples}
\end{figure*}
%%%%%%%%%%%%%%%%%

Until recently, the light curves of binary hot subdwarf systems --- unlike pulsating sdBs --- had not been subjected to thorough Fourier analyses due to the limited number of orbits for which they were observed or complicated window functions from daytime, weather and seasonal gaps. Space--based facilities like \kepler\, \ktwo, and \tess\ have changed the observational landscape, providing high signal--to--noise (S/N), uninterrupted time--series photometry for a large number of hot subdwarf binaries for the first time. 

The shapes of binary light curves are encoded by the relative amplitudes and phases of the fundamental signal and all its harmonics in a discrete Fourier transform (DFT). In fact, any coherent, periodic light curve shape can be modelled using enough terms in a Fourier series, as given by
\begin{equation}
    f(t) = f_0 + \sum^{\infty}_{i=0} a_i\sin \left(\frac{2 \pi t}{P/(i+1)} + \phi_i \right) 
\end{equation}
\noindent where $f_0$ is the mean flux, $a_i$ the amplitude of the i$^{th}$ harmonic (with $i=0$ defined as the fundamental), $\phi_i$ the phase of the i$^{th}$ harmonic, and $P$ the orbital period. The sharper the features are in the light curve (e.g., eclipses), the more Fourier terms are necessary. \cite{barlow22} recently showed that precise measurements of the relative amplitudes and phases of the harmonics can be used to classify light curves morphologies efficiently --- {\em without} directly investigating the light curves. In particular, they find that the first harmonic phase and the second harmonic strength (both with respect to the fundamental) can help distinguish between eclipsing cataclysmic variables, eclipsing sdB+dM/BD (HW Vir) binaries, non--eclipsing sdB+dM/BD binaries, and ellipsoidally modulated systems. Here we focus on applying such diagnostics to the shape of the reflection effect in eclipsing and non--eclipsing sdB+dM/BD binaries.

As shown in Figure \ref{fig:ft_examples}, the DFT of an eclipsing sdB+dM/BD binary (HW Vir) exhibits a strong fundamental signal at the orbital period and a multitude of harmonics with alternating but generally decaying amplitudes. In order to reproduce the sharp primary eclipses, the harmonics must all have minima at the same moment the fundamental does. This demands particular phase relationships between each harmonic and the fundamental. In the case of the first harmonic, we can define the phase difference between it and the fundamental as
\begin{equation}
\label{eqn:delta_phi}
    \Delta \phi_{1-0} = -2.0 \phi_0 + \phi_1
\end{equation}

\noindent Eclipsing sdB$+$dM/BD light curves exhibit a phase difference of $\Delta \phi_{1-0}$ = 90$^{\circ}$, which forces the first harmonic to have a minimum when the fundamental does and helps generates the primary eclipse.

The DFTs of non--eclipsing sdB+dM/BD binaries are much simpler. Without eclipses, only a strong fundamental and much weaker first harmonic are needed to reproduce the quasi--sinusoidal reflection effect shape, as also shown in Figure \ref{fig:ft_examples}. A phase difference of $\Delta \phi_{1-0}$ = 270$^{\circ}$ in this case aligns the maxima of the fundamental and first harmonic, thereby making the crests sharper and peaks broader. It also makes the reflection effect shape symmetric about the peak flux.

%%%%%%%%%%%%%%%%%
\begin{figure}[t]
    \centering
    \includegraphics[width=0.95\columnwidth]{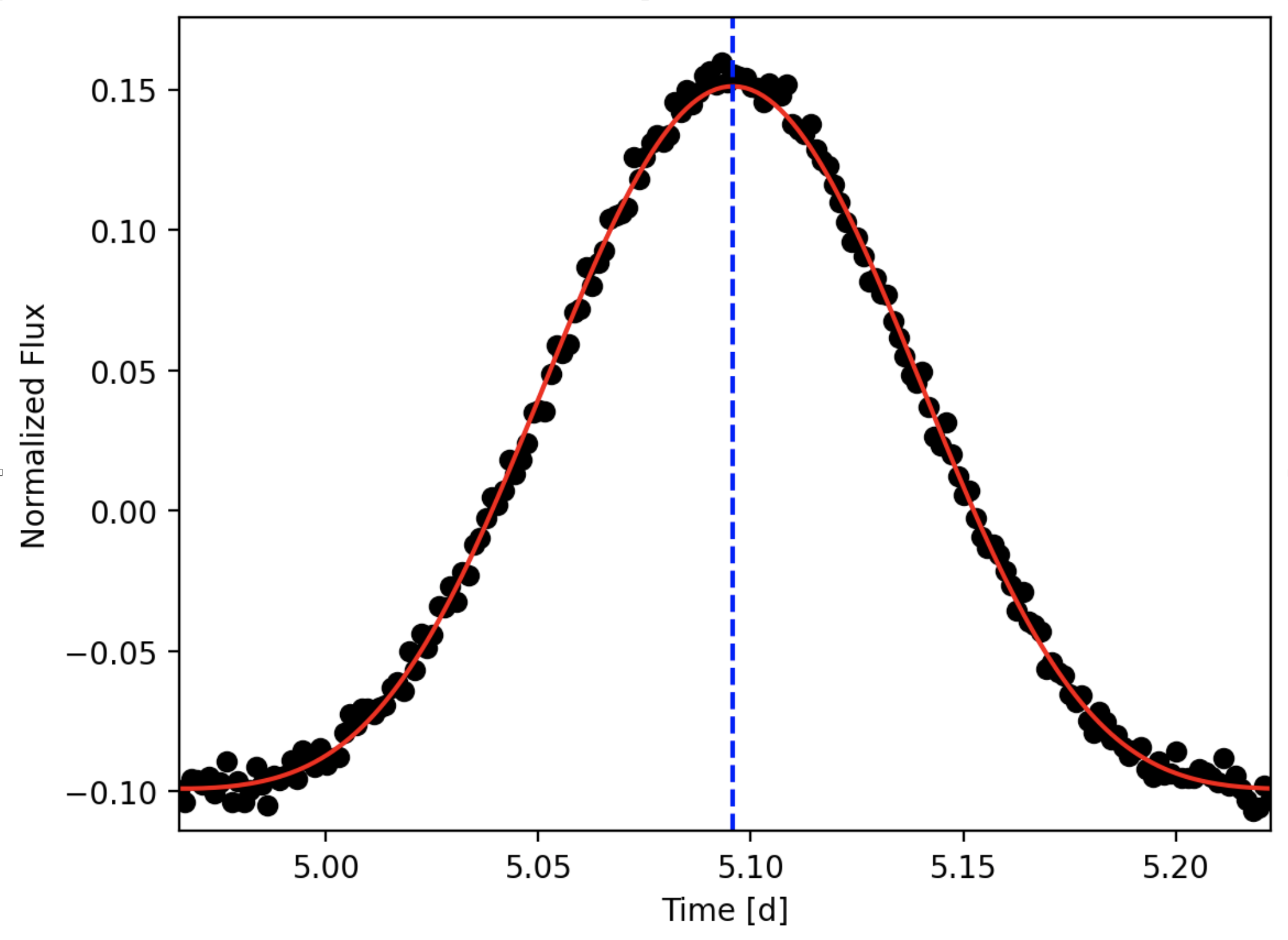}
   \includegraphics[width=0.95\columnwidth]{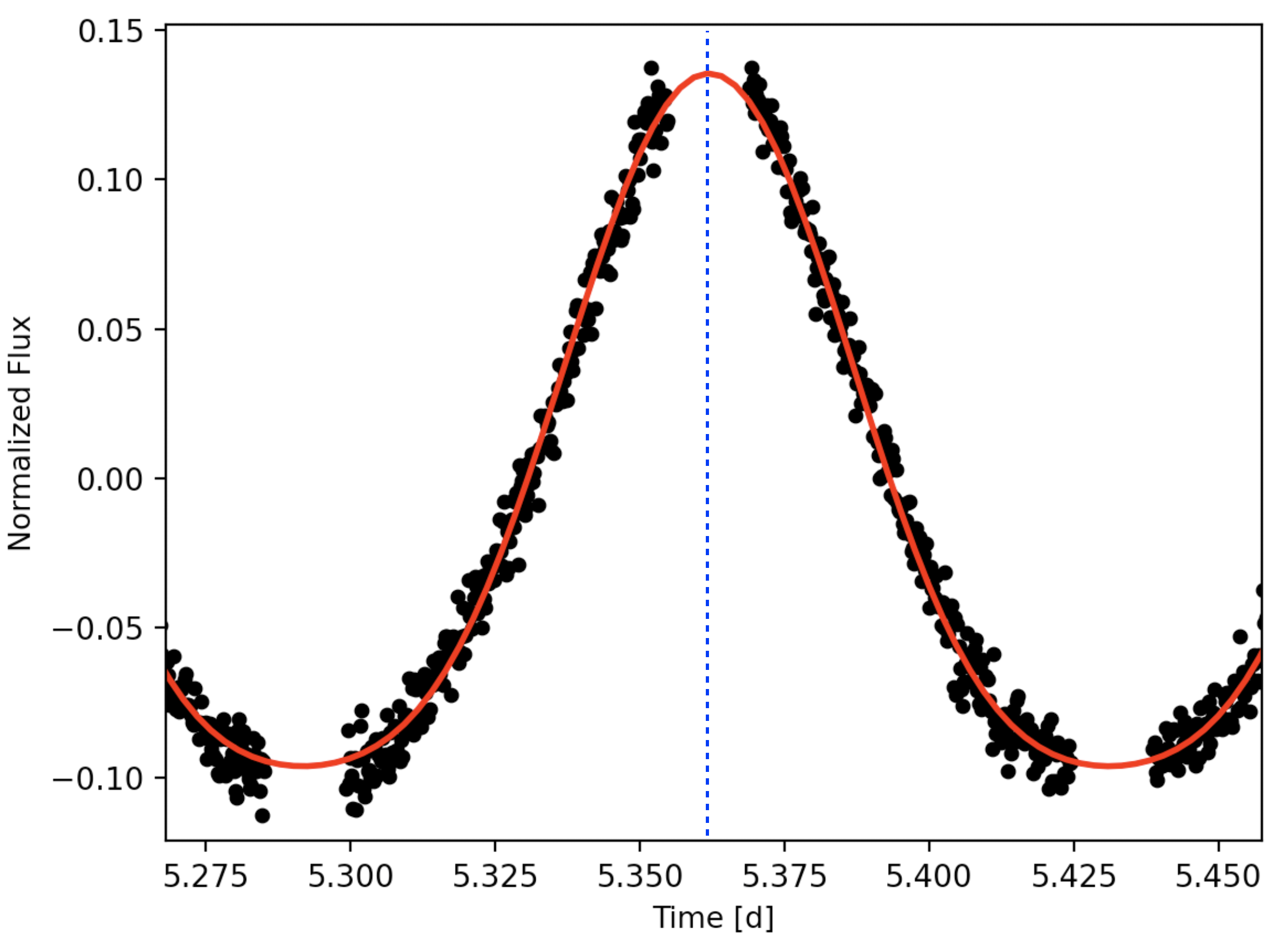}
    \caption{Example fits of Equation \ref{eqn:reflection_shape} to non--eclipsing (left) and eclipsing (right) sdB+dM/BD binaries observed by \tess. The primary and secondary eclipses have been removed from the HW Vir before fitting its reflection effect.}
    \label{fig:fit_examples}
\end{figure}
%%%%%%%%%%%%%%%%%

In order to explore potential correlations between Fourier coefficients and properties of the binaries, we used SciPy's \texttt{curve\_fit} routine to fit all observed and synthetic light curves with the summation of a fundamental and first harmonic, as given by
\begin{equation}
\label{eqn:reflection_shape}
    f(t) = f_0 + a_0 \sin \left( \frac{2 \pi t}{P} + \phi_0 \right) + a_1 \sin \left( \frac{2 \pi t }{P/2} + \phi_1 \right)
\end{equation}

%%%%%%%%%%%%%%%%%%
\begin{figure*}[h]
    \centering
    \includegraphics[width=0.95\textwidth]{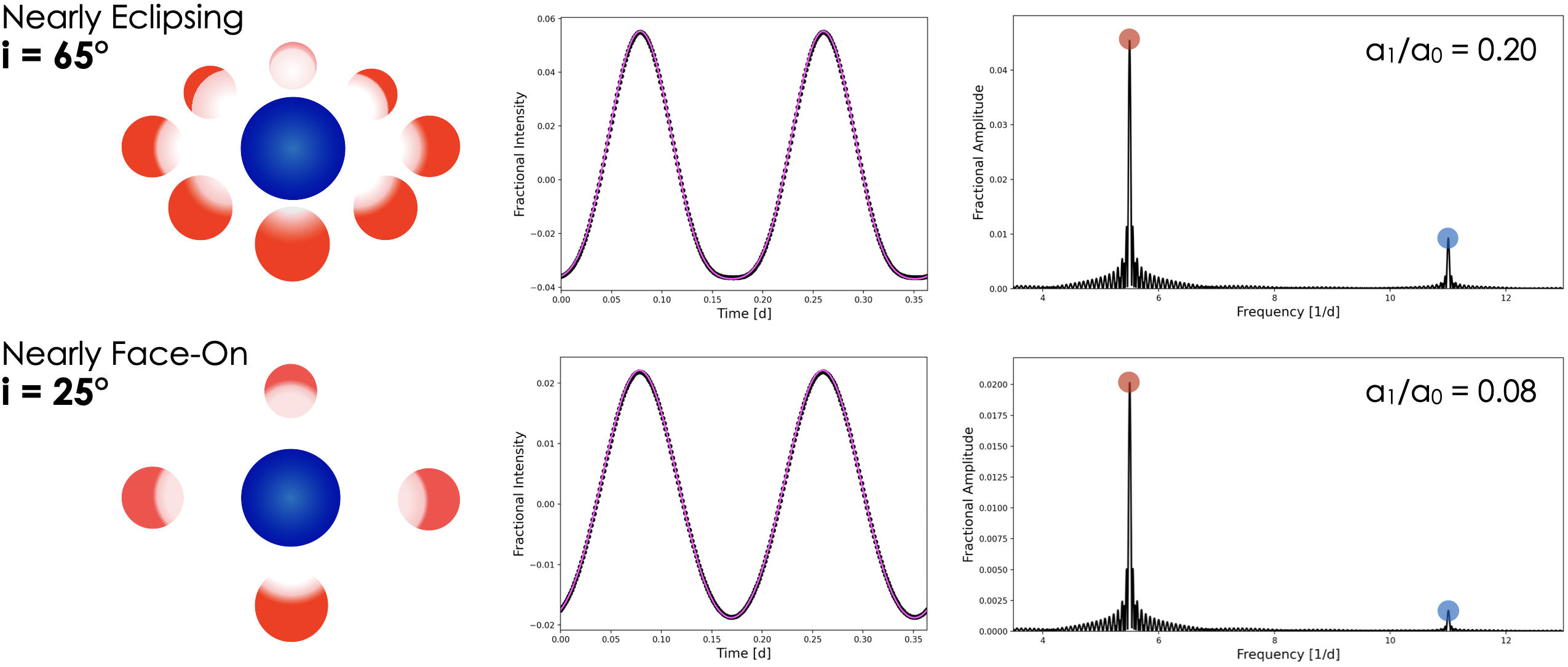} %1.75
    \caption{Illustration of the effect inclination has on reflection effect shape for high--inclination (first row) and low--inclination sdB+dM/BD binaries from \lcurve. Simple schematics representing the change in the reflection effect visbility along our line--of--sight (first column) are shown next to the resulting light curves (middle column) and their periodograms (right column). The schematic, which is presented in the reference frame of the hot subdwarf, is not to scale and for qualitative purposes only.}
    \label{fig:inclination_dft}
\end{figure*}
%%%%%%%%%%%%%%%%%%

\noindent All parameters were left free in the fit except for the first harmonic frequency, which was forced to be twice the fundamental (orbital) frequency. In the case of non--eclipsing binaries, the complete light curve was fitted. In the case of eclipsing binaries, we first cropped out the primary and secondary eclipses before fitting the light curves. This allowed us to extend our study to higher inclinations while still fitting only the reflection effect shape. Figure \ref{fig:fit_examples} presents example fits of Equation \ref{eqn:reflection_shape} to both types of systems. We recorded the best--fitting periods, amplitudes, phases,  phase differences and their uncertainties for each light curve.

%%%%%%%%%%%%%%%%%%%%%%%%%%%%%%%%%%%%%%%%%%%%%%%%%%%
%%%       INCLINATION ANGLE CONSTRAINTS         %%%
%%%%%%%%%%%%%%%%%%%%%%%%%%%%%%%%%%%%%%%%%%%%%%%%%%%
\section{Inclination from First Harmonic Strength}
\label{sec:inclination}

As previously noted, the reflection effect shape is a quasi-sinusoidal feature requiring power from both the fundamental and a first harmonic to generate the sharper crests and broader troughs of the irradiation from the cool companion. Our Fourier series fits to the synthetic light curves help us quantify this behavior. Figure \ref{fig:inclination_dft} presents example \lcurve\ models of sdB+dM/BD binaries at high and low inclinations, along with their discrete Fourier transforms and simple schematics of their orbits. The difference in curvature of the maxima and minima in the reflection effect light curves becomes more pronounced at higher inclinations. In the $i=65^{\circ}$ example shown, the first harmonic amplitude ($a_1$) needs one--fifth the amplitude of the fundamental amplitude ($a_0$) to reproduce this quasi--sinusoidal shape. As the inclination angle decreases (with orbital period fixed), the reflection effect both decreases in overall strength and becomes more sinusoidal. In the $i=25^{\circ}$ example shown, the first harmonic amplitude only needs 8\% the strength of the fundamental to reproduce the shape. 

%%%%%%%%%%%%%%%%%%
\begin{figure}[h]
    \centering
     \includegraphics[width=0.95\columnwidth]{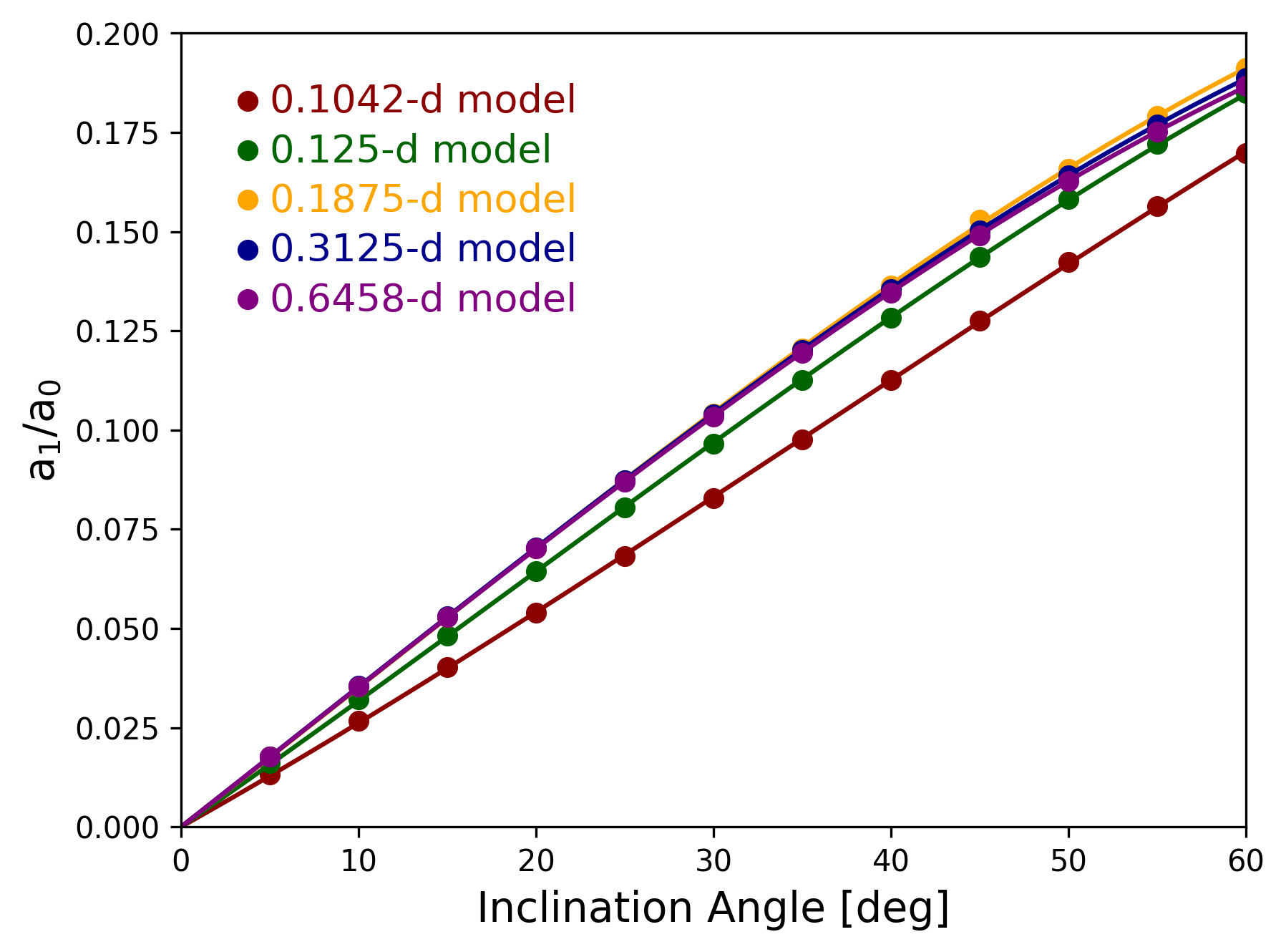} %0.95
    \includegraphics[width=0.95\columnwidth]{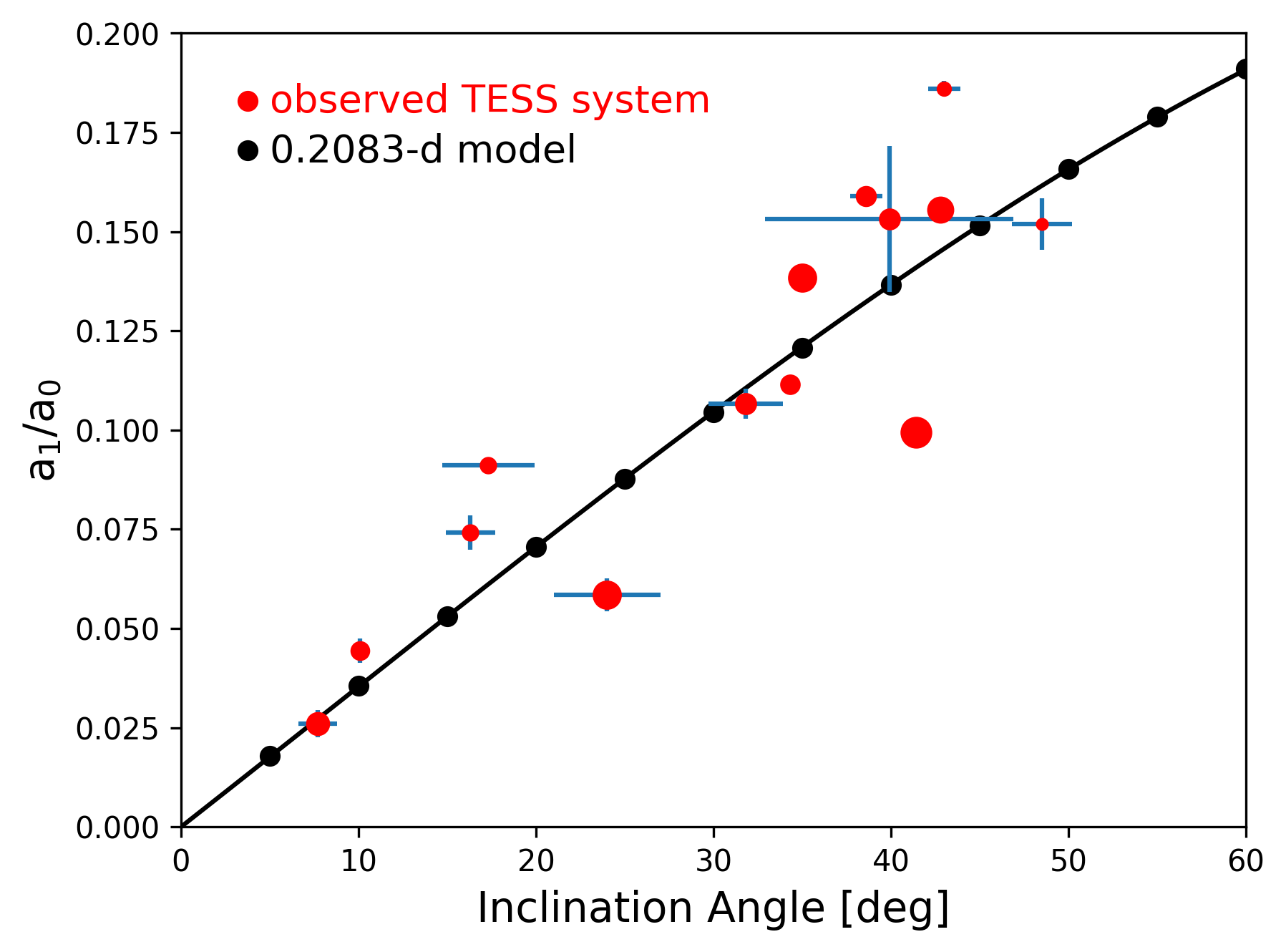} %0.95
    \caption{Reflection effect shape, as measured by the 1$^{\rm st}$ harmonic strength, plotted against inclination for non-eclipsing sdB$+$dM/BD binaries. {\em Top Panel:} Results for synthetic light curves from \lcurve\ are shown over a range of orbital periods. {\em Bottom Panel:} Measurements for fully solved reflection effect binaries from Paper I (red points) are plotted on top of results for synthetic light curves with a period of 5 hr (black points/lines), which is near the peak of the observed period distribution for sdB$+$dM/BD binaries. The sizes of the red points scale inversely with orbital period (i.e., the shortest-period systems have the largest symbols).}
    \label{fig:amp_vs_inclination}
\end{figure}

The top panel of Figure \ref{fig:amp_vs_inclination} summarizes our findings for synthetic \lcurve\ light curves of non--eclipsing binaries. For systems with a fixed period but a range of inclination angles, the first harmonic amplitude increases non--linearly with increasing inclination angle. Systems with $i<20^{\circ}$ are quite sinusoidal, with the first harmonic having less than 10\% the strength of the fundamental. The reflection effect shape for the highest--inclination systems, on the other hand, deviates quite a bit from a sinusoid, with the first harmonic strength approaching $\approx$20\% the strength of the fundamental at the transition region between non--eclipsing and eclipsing HW Vir binaries (around $i=55^{\circ}-75^{\circ}$). The orbital period also has a slight effect on $a_1/a_0$. For systems observed at the same $i$, longer--period systems have a slightly larger value of $a_1/a_0$ than shorter--period ones. This difference grows with increasing inclination. Around $i=60^{\circ}$, for example, $a_1/a_0$ can range from 0.17 to 0.19 for periods from 2 hr to 14 hrs, respectively. At a nearly face--on $i=10^{\circ}$, $a_1/a_0$ ranges from 0.025 to 0.03 for the same periods.  We note that at periods above P$\approx$0.2 d, the ratio $a_1/a_0$ actually appears to decrease slightly with increasing period at inclinations above $i=45^{\circ}$, until an asymptotic value is reached around P$\approx$0.5. We are still investigating potential causes of this effect.

For any fixed orbital period, the observed relationship between \ampratio\ and $i$ can be modelled mathematically with a third--order polynomial given by
\begin{equation}
\label{eqn:inclination}
    \frac{a_1}{a_0} = \alpha (P) \, i^3 + \beta (P) \, i^2 + \gamma (P) \,i 
\end{equation}
%\begin{equation}
%    \frac{a_1}{a_0} = \alpha \, i^3 + \beta \, i^2 + \gamma \,i + \delta
%\end{equation}
\noindent where the coefficients $\alpha(P)$, $\beta(P)$, and $\gamma(P)$ are each functions of the orbital period $P$. Since \ampratio\ must approach zero at $i$=0, there is no fourth coefficient for the degree 0 term. While the coefficients change slightly with orbital period (see Table \ref{tab:inclination}), they have quite similar values for orbital periods $P>0.2d$.

In order to test whether \ampratio\ might serve as a proxy for inclination in non-eclipsing systems, we plot in the bottom panel of Figure \ref{fig:amp_vs_inclination} all reflection effect systems with fully solved orbital parameters from Paper I on top of the results from fitting \lcurve\ models with an orbital period of 5 hr.  The latter value was chosen because it is close to the peak of the observed orbital period distribution for all non-eclipsing sdB$+$dM/BD binaries presented in Paper I, whose ranges from 0.11 d to 0.75 d. As the figure shows, the results for the well-studied, solved systems generally follow the trend of the \lcurve\ synthetic systems. The residuals of the empirical systems about the trend (along the inclination axis) have a standard deviation around 10$^{\circ}$. This striking result implies that the inclination angles of non-eclipsing reflection effect binaries can be constrained with reasonable precision from only a simple Fourier analysis of the light curve --- without the need for intensive light curve modeling with \lcurve\ or other software. We note that parameters other than orbital period and inclination angle can affect the reflection effect shape and, by extension, the values of the coefficients in Table 1. These include limb darkening vaues, temperature ratio $T_1/T_2$, and radii ratio $R_1/R_2$. Our initial investigations show that inclination has (by far) the dominant effect on \ampratio. Full exploration of the effects of the secondary parameters is beyond the scope of this paper but should be explored in due time. 

\begin{table}
\centering
\caption{Best-fitting Parameters to Equation 4 for Select \lcurve\ Synthetic Light Curves}
\begin{tabular}{cccc}
\hline
Orbital Period & $\alpha (P)$ & $\beta (P)$ & $\gamma (P)$ \\

[d] & [(1/$^{\circ}$)$^3$] &  [(1/$^{\circ}$)$^2$] &  [(1/$^{\circ}$)]\\ 
\hline
0.1042 & -1.10e-7 & 1.21e-5 & 2.51e-3\\
0.1250  & -1.55e-7 & 8.95e-6 & 3.10e-3\\
0.1875   & -1.70e-7 & 5.45e-6 & 3.47e-3\\
0.2083 & -1.56e-7 & 3.86e-6 & 3.51e-3\\
0.3125 & -1.57e-7 & 3.19e-6  & 3.52e-3\\
\hline
\end{tabular}
\label{tab:inclination}

\end{table}

%%%%%%%%%%%%%%%%%%%%%%%%%%%%%%%%%%%%%%%%%%%%%%%
%%%       REFLECTION EFFECT ASYMMETRY       %%%
%%%%%%%%%%%%%%%%%%%%%%%%%%%%%%%%%%%%%%%%%%%%%%%
\section{Beaming-Induced Reflection Effect Asymmetry}
\label{sec:asymmetry}

As previously shown, the reflection effect can be modelled mathematically with a simple Fourier series including a fundamental and first harmonic (Equation \ref{eqn:reflection_shape}). If the phase difference between the first harmonic and the fundamental is exactly \deltaphi$= 270^{\circ}$ or  $90^{\circ}$, then the resulting function $f(t)$ will be symmetric. Any departure from these two \deltaphi\ values results in an asymmetric light curve shape, assuming the first harmonic amplitude has sufficient power with respect to the fundmental.

In all post--CE sdB binaries with cool companions and periods shorter than 1.2 d, the orbits should be circular and the companions should be rotationally synchronized, given the short time scales for both processes compared to sdB lifetimes \citep{geier10}. In this case, one would expect the reflection effect to be perfectly symmetric about its flux peak in the light curve. Since the vast majority of the reflection effect shape can be reproduced using only a fundamental and first harmonic, we can use their measured phase difference to look for and quantify asymmetries in the light curves of sdB+dM/BD systems.

%%%%%%%%%%%%%
\begin{figure}
    \centering
    \includegraphics[width=0.95\columnwidth]{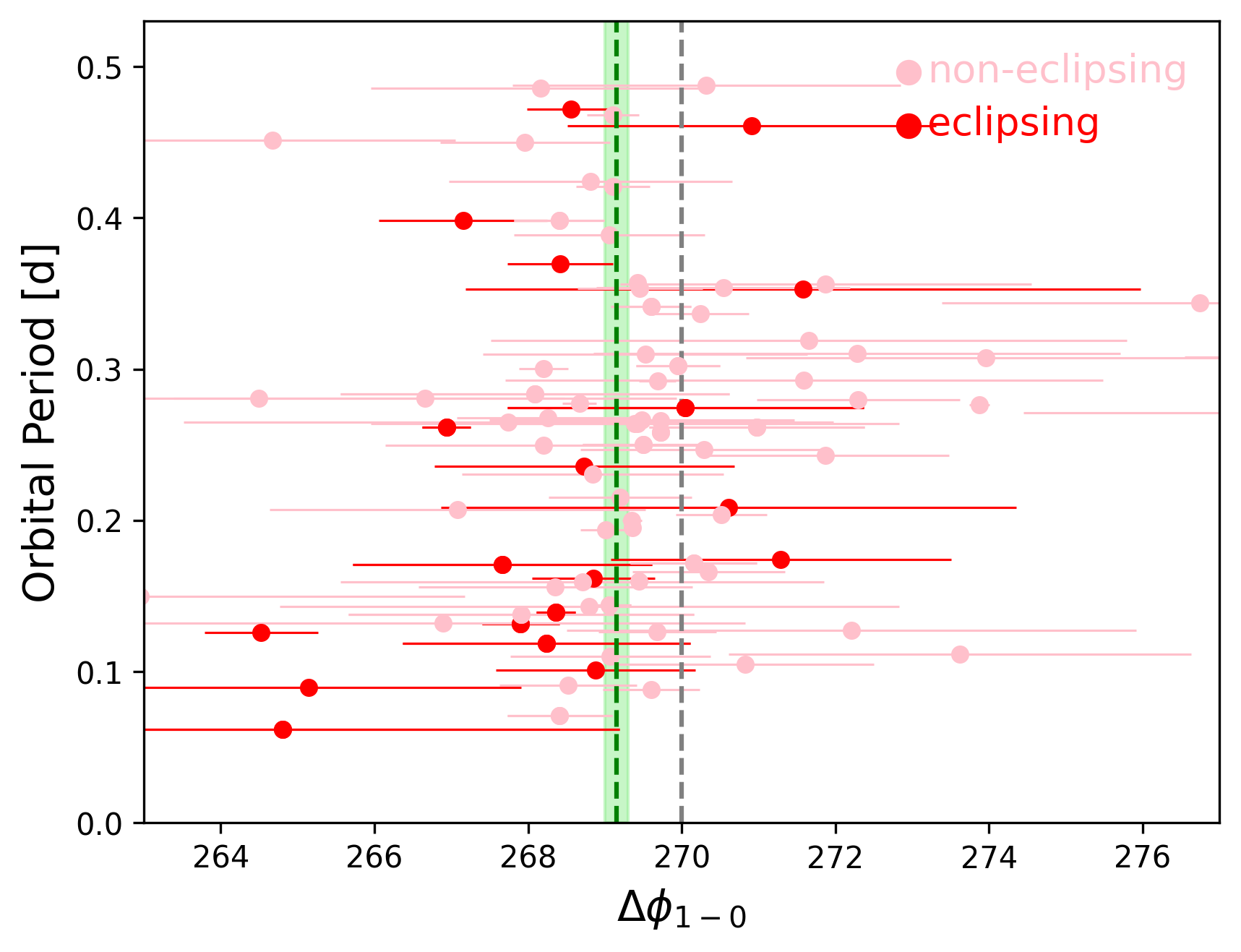} %0.95
    \includegraphics[width=0.95\columnwidth]{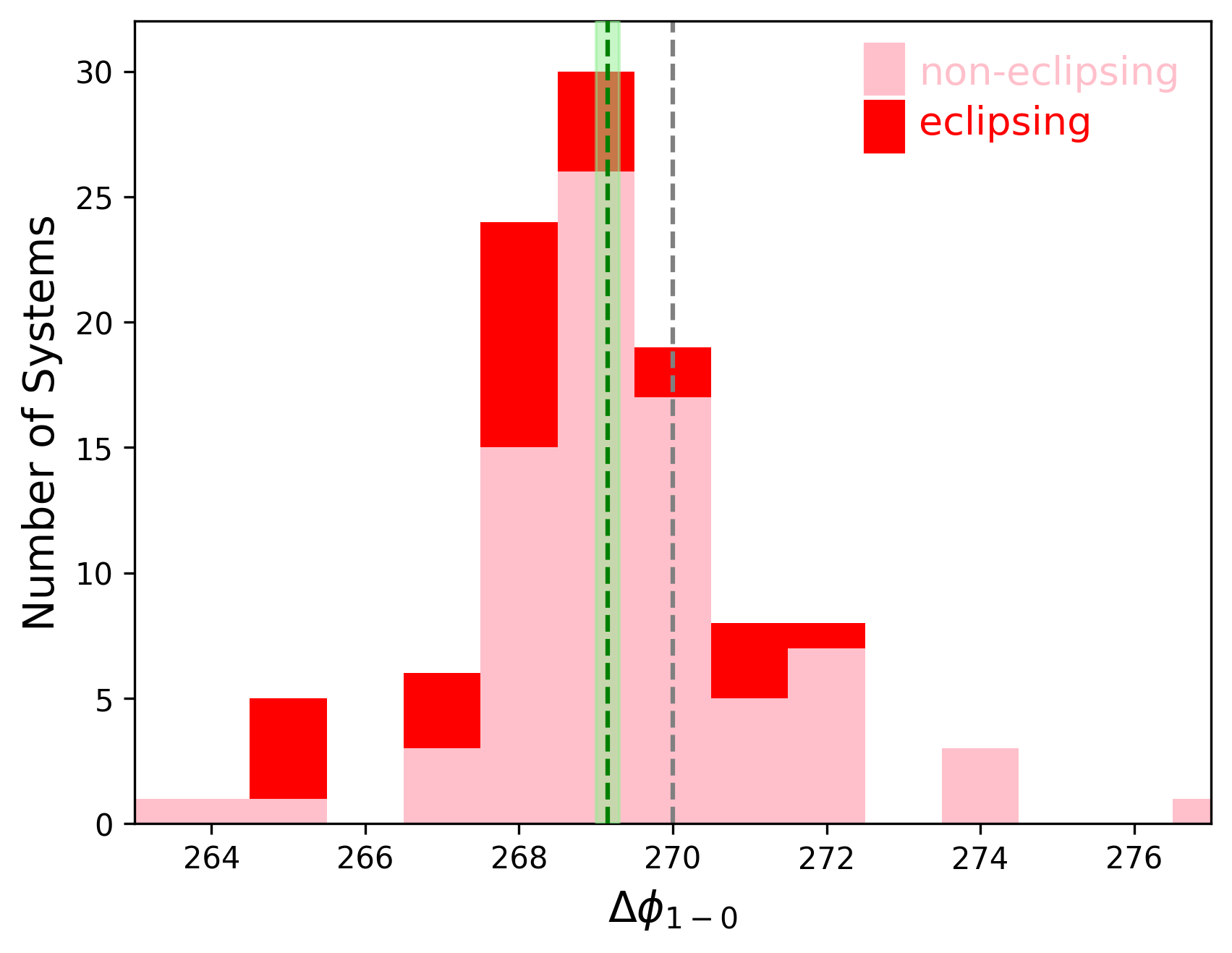} %0.95
    \caption{Reflection effect asymmetry in sdB+dM/BD binaries, as measured by the phase difference between the 1$^{\rm st}$ harmonic and fundmanental from Fourier series fits to \tess\ light curves. {\em Top Panel:} Phase differences are shown plotted against orbital period for individual eclipsing (red; HW Vir) and non-eclipsing (pink) binaries. {\em Bottom Panel:} Histogram of the measured phase differences with the same color coding. The gray lines mark 270$^{\circ}$, which would result in a symmetric reflection effect. The green lines and shaded regions denote the peak of the empiricial distribution and its uncertainty.}
    \label{fig:phases_vs_period}
\end{figure}
%%%%%%%%%%%%%

Figure \ref{fig:phases_vs_period} presents empirical \deltaphi\ values from fitting Equation \ref{eqn:reflection_shape} to our \tess\ light curves, plotted against orbital period. A histogram of the results is also shown. We find that both non--eclipsing and eclipsing systems show \deltaphi\ distributions that are centered {\em below} $270^{\circ}$ and in rough agreement with one another. Consequenty, the reflection effect shape in sdB+dM/BD binaries appears to be asymmetric. We find a median phase difference of \deltaphi$=269.15 \pm 0.15 ^{\circ}$ when combining measurements from both eclipsing and non-eclipsing systems. With \deltaphi $< 270^{\circ}$, our \tess\ light curves are skewed such that they reach their highest flux values {\em after} the halfway point between flux minima. They rise more slowly from minimum to maximum flux and decay more rapidly from maximum to minimum. 

%%%%%%%%%%%%%%%%%
\begin{figure}[t]
    \centering
    \includegraphics[width=0.95\columnwidth]{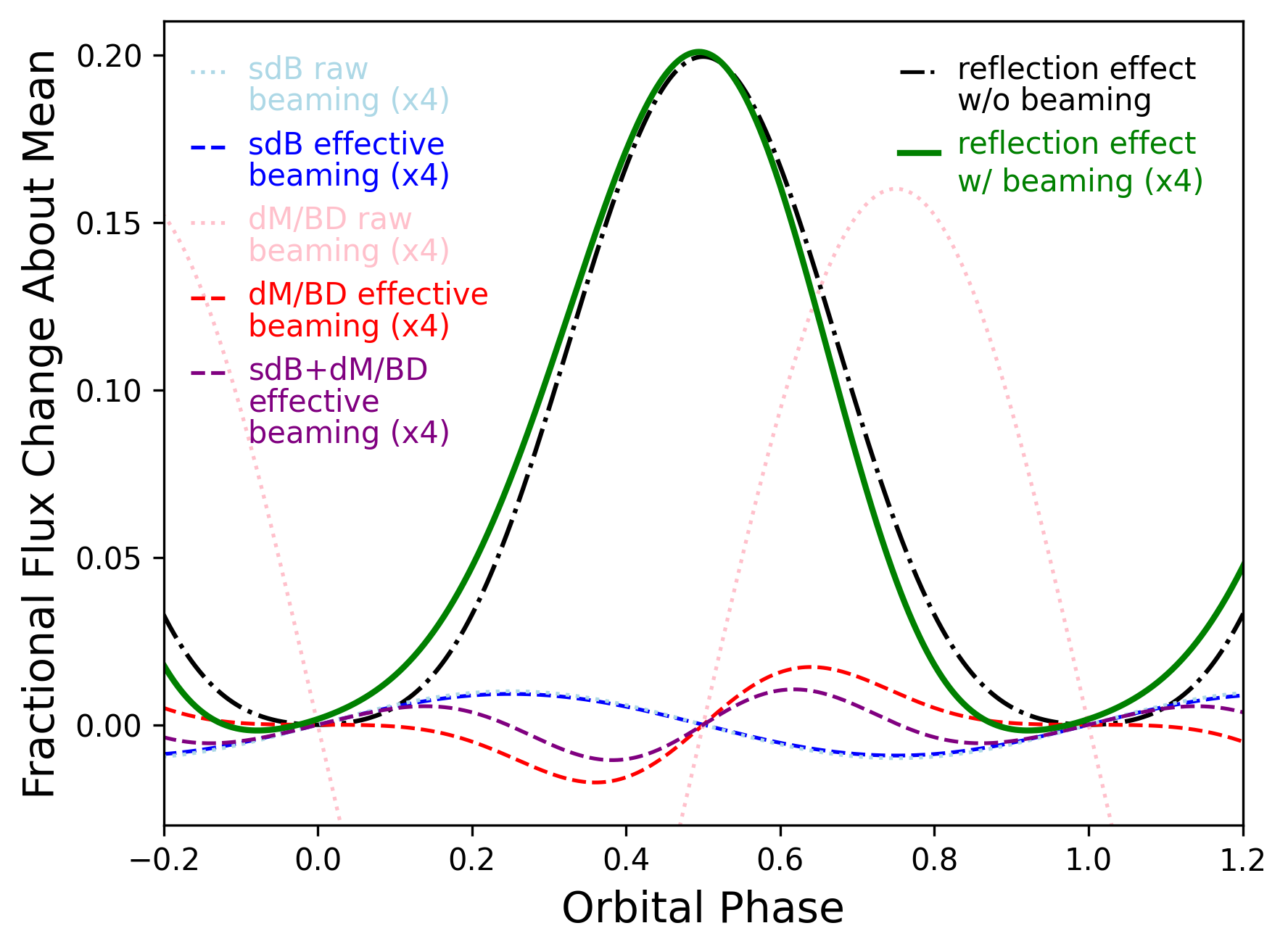} %0.95
    \caption{Illustration of the reflection effect asymmetry induced by (highly exaggerated) relativistic beaming (solid green line). The reflection effect would be symmetric in the absence of beaming (dot-dashed black line), assuming a synchronized/circularized binary. The beaming effect from the sdB (dashed blue line) is sinusoidal and peaks at phase 0.25. The beaming effect from the dM/BD, whose velocity variations are $\pi$ out-of-phase with those of the sdB, {\em would} be sinusoidal if the cool companion contributed non-negligible and constant flux to the system light (dotted pink line). However, the dM/BD contributes negligible light to the system outside its reflection effct, and its beaming shape is more complicated (dashed red line) since it can only manifest itself through the reflection effect. Both beaming effects, which are shown approximately four times their typical strength for improved visualization purposes, lead to an overall skewness of the observed reflection effect.} 
    \label{fig:reflection_sinusoids}
\end{figure}

We hypothesize that the observed asymmetry arises from relativistic beaming from both stars, which we illustrate in Figure \ref{fig:reflection_sinusoids}. The beaming variation from the sdB, assuming it is otherwise constant in flux, should be sinusoidal, peak at phase 0.25 (maximum blueshift of sdB), and have amplitudes around 0.25\% in typical sdB+dM/BD binaries (blue line in Fig. \ref{fig:reflection_sinusoids}). Although the cool companion gives off only a negligible fraction of the system light, it also contributes a beaming variation through the reflection  effect, which can contribute up to 20\% to the total flux at orbital phase 0.5. If the contribution from the companion to the total system light were non--negligible and constant in time, its beaming signal would also take the form of a simple sinusoidal oscillation with a peak at phase 0.75 (maximum blueshift of dM/BD) and trough at phase 0.25 (pink line in Fig. \ref{fig:reflection_sinusoids}). In reality, however, the observed pattern is more complicated since the beaming factor only operates on the reflection effect itself, which constantly changes in strength throughout the orbit. The resulting beaming variation is a quasi-sinusoidal oscillation with both trough and crest compresssed near phase 0.5 (red line in Fig. \ref{fig:reflection_sinusoids}). Adding both the sdB and dM/BD beaming effects together moves the observed phases of both the flux minimum and maximum to slightly earlier phases (purple line in Fig. \ref{fig:reflection_sinusoids}). Because the sdB and dM/BD beaming effects cancel out more around phase 0.5 than they do around phase 0.0/1.0, the observed phase of flux minimum is pushed further to earlier phases than the flux maximum is, and the net effect is an overall skewing of the reflection effect shape (green line in Fig. \ref{fig:reflection_sinusoids}).

%%%%%%%%%%%%%%%%%
\begin{figure}[t]
    \centering
    \includegraphics[width=0.95\columnwidth]{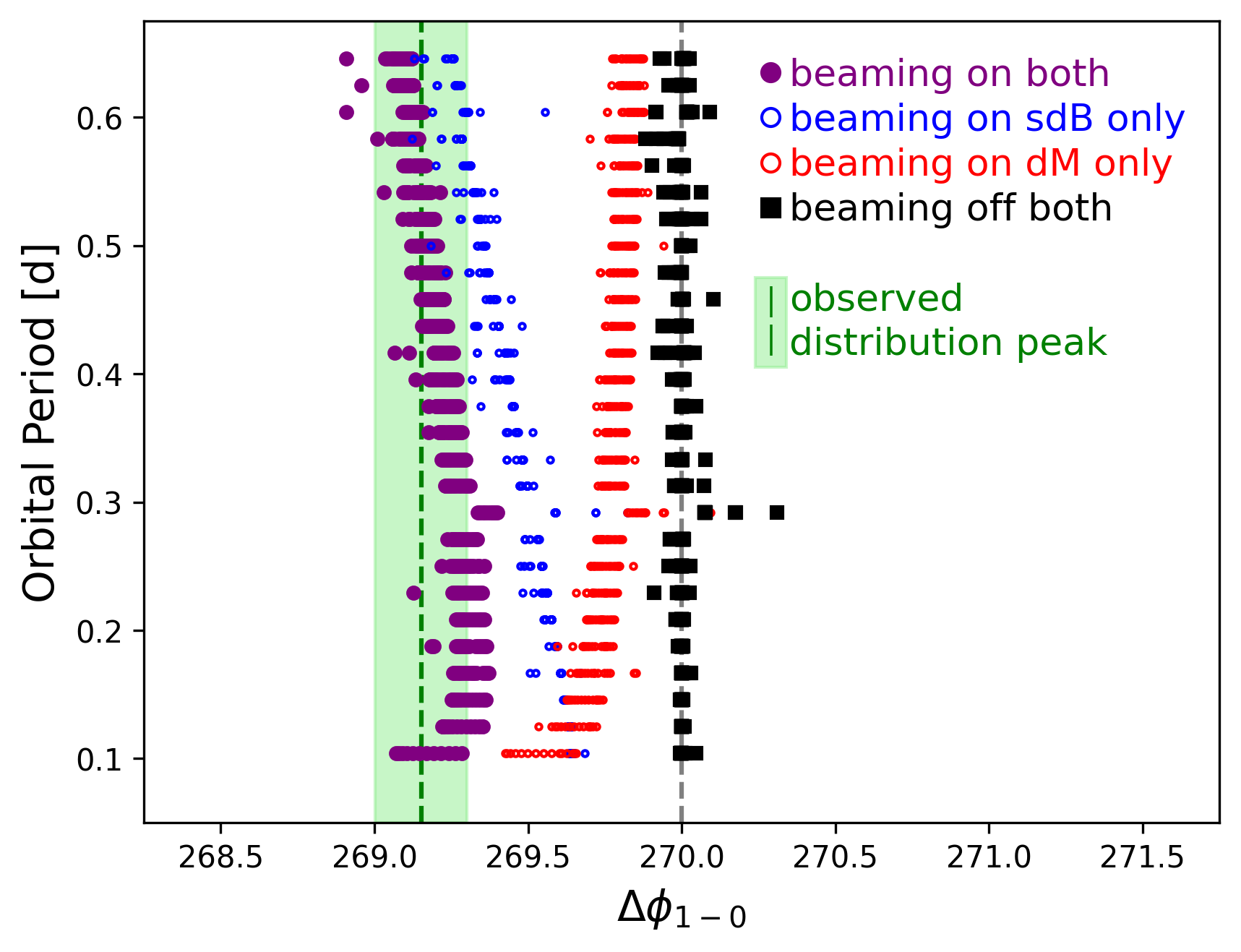} %0.95
    \caption{Same as the top panel of Figure  \ref{fig:phases_vs_period}, but for synthetic data from \lcurve. Results are shown for four sets of light curves: relativistic beaming turned off for both stars (black squares), beaming turned on for the cool companion only (red open circles), beaming turned on for the hot subdwarf only (blue open circles), and beaming turned on for both stars (purple filled circles). The dashed gray line marks 270$^{\circ}$ (symmetric shape) while the dashed green line and shaded region mark the peak of the empirical \deltaphi\ distribution and its uncertainty.} 
    \label{fig:phases_models}
\end{figure}

In order to test the hypothesis that the observed \deltaphi\ offset and reflection effect asymmetry arise from Doppler beaming, we ran the same Fourier series analysis on the synthetic light curves from \lcurve\ we did on our \tess\ data. For each model system, four separate sets of light curves were generated: (i) beaming turned on for both stars; (ii) beaming turned off for both stars; (iii) beaming turned on for the sdB only; and (iv) beaming turned on for the dM/BD only. Figure \ref{fig:phases_models} presents the resulting phase measurements.

As with the empirical light curves, we detect a clear asymmetry in the reflection effect of model light curves from \lcurve, when beaming is turned on for both stars (purple circles in Fig. \ref{fig:phases_models}). The model \deltaphi\ values generally agree with the \tess\ light curve measurements in both direction and magnitude. At the shortest periods, they cluster around \deltaphi\ $= 269.1^{\circ}$, and as the period increases, they generally decrease in value. Upon running the same analysis on the same \lcurve\ systems but with beaming turned {\em off} for both stars (black squares), we no longer measure an asymmetry and find \deltaphi\ $\approx$ 270$^{\circ}$. This exercise demonstrates that the majority of the measured reflection asymmetry --- if not {\em all} of it --- arises from relativistic beaming. The scatter in the points seen at each fixed orbital period primarily arises from the range of inclination angles used in the models.

Further inspection of Figure \ref{fig:phases_models} reveals the beaming contributions from each star to the observed asymmetry. When beaming is turned on {\em only} for the cool companion (red circles in Fig. \ref{fig:phases_models}), the \deltaphi\ values are smallest (around 269.4$^{\circ}$) at the shortest periods and asymptotically approach \deltaphi\ $= 270^{\circ}$ as the period increases. At the shortest periods, the reflection effect is strongest and the velocities are highest. These two effects work together to make the asymmetry more pronounced. As the orbital period increases, the strength of the reflection effect decreases, as does the orbital velocity. Consequently, both the fractional contribution of the reflection effect to the total system flux and the beaming strength decreases, and the reflection effect shape becomes more symmetric. 

Turning the beaming on for only the sdB (blue circles in Fig. \ref{fig:phases_models}) shows it also contributes to the asymmetry. Despite its corresponding velocity variation being $\pi$ out of phase with the cool companion, the hot subdwarf beaming skews the reflection effect shape in the same direction, resulting in \deltaphi\ $< 270^{\circ}$. At the shortest orbital periods, sdB beaming induces an asymmetry with \deltaphi\ $\approx 269.5^{\circ}$. As the period increases, \deltaphi\ actually {\em decreases} in value, implying the asymmetry grows. This can be explained by the strength of the reflection effect quickly dropping off and the beaming variations from the sdB, which peak at phase 0.25, becoming increasingly dominant. As this happens, the overall flux peak in the light curve shifts more and more from phase 0.5 to phase 0.25, the fundamental in the Fourier series follows it, and the \deltaphi\ phase difference between the fundamental and first harmonic decreases. It should asymptotically reach \deltaphi\ $= 180^{\circ}$, but the rapid decay in the strength of the reflection effect and, more importantly, sdB Doppler beaming at periods greater than 1 d make measurements in this regime nearly impossible.

With beaming turned on for both stars (purple circles in Fig. \ref{fig:phases_models}), the two effects discussed above work together to create the overall asymmetry we observe in the reflection effect of sdB+dM/BD binaries. At the shortest orbital periods, the velocities of both stars are highest, and thus effects due to beaming are more prevalent. However, the reflection effect also being strongest at shortest periods allows the cool compainon's beaming effect to carry more weight than at longer periods. At $P \approx 2 hr$ the asymmetry is quite strong with \deltaphi\ $\approx$ 269$^{\circ}$. As the orbital period increases from 2 hr to 4 hr, \deltaphi\ actually increases, and the reflection effect shape becomes slightly more symmetric. This is due to the rapid drop off in both the reflection effect amplitude and cool companion velocity. As periods increase beyond 4 hr, the beaming contribution from the hot subdwarf 
 begins to overwhelm the rapidly decaying reflection effect, and the the \deltaphi\ values begins their slow decline.

%%%%%%%%%%%%%%%%%%%%%%%%%%
%%%       SUMMARY      %%%
%%%%%%%%%%%%%%%%%%%%%%%%%%
\section{Summary}
\label{sec:summary}

In Papers I \& II, we solved for the system parameters of several new and previously known sdB+dM/BD binaries using high S/N light curves from \tess. One remarkable result from this work was our ability to constrain the inclination angle and solve for the masses in {\em non-eclipsing} systems. Until then, this had only been accomplished for a handful of eclipsing (HW Vir) systems. Originally driven by a desire to understand why the inclination could be constrained so precisely in these non-eclipsing systems, we computed in this paper the discrete Fourier transforms of all solved sdB+dM/BD binaries with \tess\ light curves and investigated correlations between their system parameters and the relative amplitudes and phases of the first harmonic and fundamental.

The inclination angle strongly affects not only the amplitude of the reflection effect, but also its overall {\em shape}. Systems with higher inclinations have sharper crests and broader troughs, and those that are nearly face-on have much more sinusoidal reflection effect shapes. From an analysis of synthetic light curves, we find that this effect is encoded by the amplitude ratio of the first harmonic to the fundamental (Figure \ref{fig:amp_vs_inclination}) in the DFT of the light curve. Differences in the shapes are relatively subtle, but they are clearly detectable in \tess\ photometry due to its high S/N and long baselines. This helps explain why the light curve modeling in Papers I \& II led to such robust results for non-eclipsing systems. Figure \ref{fig:amp_vs_inclination} implies that one can constrain the inclination angle in non-eclipsing sdB+dM/BD binaries to within $\approx$10$^{\circ}$ deg from their light curve DFTs alone. Although intense light curve modeling of individual systems with codes like \texttt{lcurve} will generally produce the best results, this process can be computationally expensive and time-consuming for a large number of targets. Using the first harmonic ratio to estimate the inclination angle could be advantageous when dealing with large numbers of binaries from current and future photometric surveys (e.g., Argus, LSST, etc.). One could make quick and efficient work of assessing the distribution of inclination angles for many reflection effect binaries, analyzing them all in a consistent way to avoid systematic offsets between different modeling codes. Moreover, the first harmonic-to-fundamental ratio is much less susceptible to pixel contamination from nearby stars, which can be common in photometric surveys with large fields of view. Contamination can reduce the observed amplitudes in reflection effect binaries without significantly affecting the ratio of harmonic amplitudes.

In our work investigating DFTs, we discovered an asymmetry in the reflection effect that we attribute to Doppler beaming (Figure \ref{fig:phases_vs_period}). %The second half of the orbit appears to be slightly brighter than the first half despite the geometric symmetry present in such systems (Figure \ref{fig:phases_vs_period}). 
An analysis of synthetic light curves shows that beaming from both the sdB {\em and} the dM/BD companion contribute to this skewness, and in the same direction (Figure \ref{fig:phases_models}). %The beaming signal from the cool companion, acting only through the reflection effect, adds to the flux signal between phases 0.5 and 1.0 and subtracts from it between phases 0.0 and 0.5. Although its corresponding velocity signal is $\pi$ out of phase with that of the companion, beaming from the sdB adds to the skewness in the same direction. After reflection effect peak, when the sdB moves away from Earth, its observed flux decreases due to beaming. Before reflection effect peak, the opposite occurs. 
The net result is a reflection effect light curve in which the flux increases from minimum to maximum more slowly than it decays from maximum to minimum (Figure \ref{fig:reflection_sinusoids}). We note that we were unable to detect beaming in Papers I \& II through direct \lcurve\ modeling of the phase-folded light curves, and this is likely due to two reasons. First, the fractional flux variations it induces in reflection effect binaries are quite small --- on the order of a few ppt. Second, \lcurve\ struggles to pefectly model the irradiation effect, especially for the highest-inclination systems. As shown in Figure D.1 of Paper II, \lcurve\ models overestimate the flux at the peak of the reflection effect but underestimate it immediately before and after. The model shortcomings lead to residuals on the order of 0.25\%-0.5\% that are larger than the flux variations expected from Doppler beaming. Thus, until we improve the treatment of the irradiation effect in \lcurve, we will be unable to model beaming in the phase-folded light curves reliably.

Overall, our results indicate that the inclination angle and the radial velocities of {\em both} stars are encoded within the light curves of eclipsing and non-eclipsing sdB+dM/BD binaries, if the S/N is high enough. Since non-eclipsing binaries outnumber HW Virs by a factor of 2-3, our ability to constrain their inclination angles opens up many more systems to advanced modeling than in the past, especially as current and future photometric surveys gather additional high S/N light curves. Moreover, the ability to disentangle both stellar velocities would represent a true paradigm shift in the study of sdB+dM/BD binaries, which are in general single-lined binaries due to their extreme luminosity ratios. 

In order to fully realize the potential for high S/N light curves to provide precise inclinations and radial velocities, one must first understand and quantify numerous second-order effects. In the case of the reflection effect shape, limb darkening, gravity darkening, the orbital period, the ratio of stellar radii, and the ratio of surface temperatures all have small secondary effects on \ampratio\ that must be taken into account before the inclination angle can be derived from the reflection effect shape. The same is true for \deltaphi\ and Doppler beaming.  For example, the R\o mer delay should lead to the reflection effect peak arriving a few seconds later than the halfway point between flux minima (see, e.g., \citealt{bar12}). The resulting temporal compression of the second half of the orbit results in another asymmetry, albeit a miniscule one, in the same direction as that generated by beaming. Our \lcurve\ models and results also assume complete circularization and tidal synchronization of the orbit. Although this should be the case for most of our post-CE binaries \citep{geier10}, additional asymmetries would be introduced into the light curve and have to be removed if either assumption proves false for a particular system. The relatively small standard deviation of \deltaphi\ measurements surrounding the expected asymmetry value from beaming alone implies that these secondary effects cannot be that large. Full investigation of all of the above effects is beyond the scope of the current paper. Nonetheless, our exploratory work reveals that high S/N light curves of both eclpising and non-eclipsing reflection effect sdB+dM/BD binaries hold more information than previously realized. As the age of large photometric surveys with moderate-to-high cadence continues, DFT analyses may permit some fundamental properties of an ensemble of sdB binaries to be determined quickly and efficiently.

\begin{acknowledgements}
B.B., I.P. and B.S. acknowledge financial support from NASA Grant 80NSSC21K0364 and National Science Foundation award AST \#1812874. B.B. and B.S. would also like to acknowledge High Point University for support of parts of this project through their 2022 Summer Research Program in the Sciences. V.S. acknowledges funding from the German Academic Exchange Service (DAAD PPP USA 57444366) and the Deutsche Forschungsgemeinschaft under grant GE2506/9-1. T.K. acknowledges support from the National Science Foundation through grant AST \#2107982, from NASA through grant 80NSSC22K0338 and from STScI through grant HST-GO-16659.002-A. Co-funded by the European Union (ERC, CompactBINARIES, 101078773). Views and opinions expressed are however those of the author(s) only and do not necessarily reflect those of the European Union or the European Research Council. Neither the European Union nor the granting authority can be held responsible for them.
     
This research made use of Lightkurve, a Python package for Kepler and \textit{TESS} data analysis \citep[][]{lightkurve18}. This paper includes data collected by the \textit{TESS} mission, which are publicly available from the Mikulski Archive for Space Telescopes (MAST). Funding for the \textit{TESS} mission is provided by NASA's Science Mission directorate.
\end{acknowledgements}

\bibliography{biblio}
\bibliographystyle{aa}

%\begin{appendix}
%\newpage\onecolumn
%\section{Parameters of the close sdB binaries in TESS}
%\begin{longtable}{llllllll}
%\caption{Atmospheric and absolute parameters of the sdB binaries with spectroscopic parameters and with space-based light curves determined by spectroscopy \citep[][and references therein and references in Table \ref{refl}]{Kupfer2015} and spectral energy distribution fitting together with the \textit{Gaia} parallax.}

%\label{sed_known}\hline
%target & $T_{\rm eff,spec}$ & $\log{g}_{\rm spec}$ & $T_{\rm eff,sed}$  & $M_{\rm sed}$ & $L_{\rm sed}$ & $R_{\rm sed}$ \\
%& [K ] & [cgs] &  [K]  & [M$_\odot$] & [L$_\odot$] & [R$_\odot$] %\\\hline
%\multicolumn{8}{c}{Reflection effect systems}\\\hline
%KPD2215+5037 & 29600 $\pm$ 1000& 5.64$\pm$ 0.10 & %$27000^{+7000}_{-6000}$ & $0.445^{+0.063}_{-0.055}$ & %$19.9^{+1.9}_{-1.7}$ & $0.170^{+0.005}_{-0.005}$ \\

%\hline
%\end{longtable}

%$^a$ paper II %\qquad $^b$ \citet{baran19} \qquad $^c$ \citet{nemeth:2012,kawka:2015}

% WARNING
%-------------------------------------------------------------------
% Please note that we have included the references to the file aa.dem in
% order to compile it, but we ask you to:
%
% - use BibTeX with the regular commands:
%   \bibliographystyle{aa} % style aa.bst
%   \bibliography{Yourfile} % your references Yourfile.bib
%
% - join the .bib files when you upload your source files
%-------------------------------------------------------------------

\end{document}